# PREDICTIVE COMPARATIVE QSAR ANALYSIS OF AS 5-NITROFURAN-2-YL DERIVATIVES MYCO BACTERIUM TUBERCULOSIS H37RV INHIBITORS

Doreswamy[1] and Chanabasayya .M. Vastrad[2]

[1]Department of Computer Science Mangalore University , Mangalagangotri-574 199, Karnataka, INDIA

[2]Department of Computer Science Mangalore University , Mangalagangotri-574 199, Karnataka, INDIA

## ABSTRACT

*Antitubercular activity of 5-nitrofuran-2-yl Derivatives series were subjected to Quantitative Structure Activity Relationship (QSAR) Analysis with an effort to derive and understand a correlation between the biological activity as response variable and different molecular descriptors as independent variables. QSAR models are built using 40 molecular descriptor dataset. Different statistical regression expressions were got using Partial Least Squares (PLS) ,Multiple Linear Regression (MLR) and Principal Component Regression (PCR) techniques. The among these technique, Partial Least Square Regression (PLS) technique has shown very promising result as compared to MLR technique A QSAR model was build by a training set of 30 molecules with correlation coefficient $(r^2)$ of 0.8484 , significant cross validated correlation coefficient $(q^2)$ is 0.0939, F test is 48.5187, $r^2$ for external test set $(pred\_r^{2})$ is -0.5604, coefficient of correlation of predicted data set $(pred\_r^2se)$ is 0.7252 and degree of freedom is 26 by Partial Least Squares Regression technique.*

## 1. INTRODUCTION

Tuberculosis in humans is generally caused by mycobacterium tuberculosis(TB). The desease is spread by respirable droplets generated during effective expiratory manoeuvres such as coughing. TB desease can be either active or latent[1] . The World Health Organization (WHO) asses that within the next twenty years about thirty million people will be troubled with the bacillus [2-3]. The analytic management of TB has depends dully on a limited number of drugs such as Isonicotinic acid, Hydrazide,Rifadin, Rimactane ,Myambutol ,Streptomycin, Ethionamide, Pyrazinamide, Fluroquinolones etc [4]. Still with the origin of these special chemical drugs the spread of TB has not been eradicated completely because of delayed treatment programmes .There is now recognition that new drugs to treat TB are necessarily required, particularly for use in shorter medication procedure than are possible with the current agents and which can be engaged to treat multi-drug resistant and hidden disease[5].

5-nitrofuran-2-yl  shows effective in vitro and in vivo antimycobacterial activity [6]. There is also a great effort to find and develop newer, 5-nitrofuran-2-yl, and some of them might have value in the remedy of TB [7].  Chemo informatics[26]   and computer-aided drug design (CADD) are likely to contribute to a possible solution for the dangerous situation regarding this infectious disease by  assisting in the swift identification of new effective anti-TB agents. The other way for overcoming the absence of empirical analysis for biological systems is depends on the activity to develop quantitative structure activity relationship (QSAR) [8] . QSAR models are mathematical expressions formulating a relationship between chemical structures and biological activities. These models have different capability, which is providing a deeper knowledge about the process of biological activity. In the first step of a usual QSAR study one needs to find a set of molecular descriptors with the higher influence on the biological activity of interest [9]. A broad scope of molecular descriptors[10] has been used in QSAR modeling. These molecular descriptors[11] have been categorised  into different classes, including constitutional, geometrical, topological, quantum chemical and so on. Using  such an way one could predict the activities of newly formulated compounds before a conclusion is being made whether these compounds should be truly synthesized and  tested.  We examine the performance of  Partial Least Squares(PLS) based QSAR models with the results produced by Multi Linear Regression(MLR ) and Principal Component Regression (PCR) methods to discover basic requirements for additional bettered antitubercular activity.

## 2. MATERIALS AND METHODS

### 2.1 MOLECULAR DESCRIPTOR  DATA SETS

A set of fourty molecule compounds relates to derivatives for mycobacterium TB(H37Rv) inhibitors were taken from large antitubercular drug molecule databases[12] using substructure mining tool Schrodinger Canvas 2010(Trial version) [13]. All molecules were handled by the Vlife MDS [14]  - 2D coordinates of atoms were recalculated counter ions and salts were eliminated from molecular structures, molecules were neutralized, mesomerized and aromatized. Data sets were then refined from duplicates. The 2D-QSAR models were produced using a training set of thirty molecules. Predictive ability of the models was assessed by a test set of ten molecules   with consistently distributed biological activities. The observed selection of test set molecules was made by seeing the fact that test set molecules shows a range of biological activity similar to the training set. The actual  and predicted biological activities  of the training and test set molecules are given in Table 1.

| Sl no | Compound | IC50a(µg/ml) | PIC50b | | Residual |
|-------|----------|--------------|--------|------|----------|
| | | | Obsr | Pred | |
| 1 | 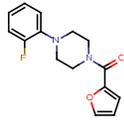 | 4.41 | 5.355 | 5.3820 | 0.027 |
| 2 | 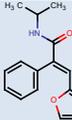 | 6.41 | 5.193 | 5.2282 | 0.0352 |
| 3 | 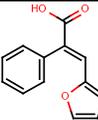 | 3.29 | 5.482 | 5.2282 | 0.2538 |
| 4 | 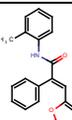 | 8.7 | 5.060 | 5.2282 | 0.1682 |
| 5 | 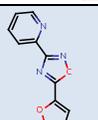 | 11.38 | 4.943 | 6.0161 | 1.0731 |
| 6 | 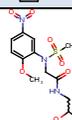 | 6.83 | 5.165 | 5.3820 | 0.217 |
| 7 | 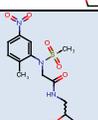 | 2.53 | 5.596 | 5.3820 | 0.214 |
| 8 | 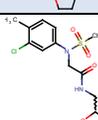 | -4.35 | 5.361 | 5.3820 | 0.021 |
| 9 | 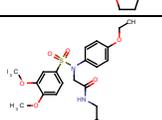 | -0.95 | 6.022 | 5.3820 | 0.64 |
| 10 | 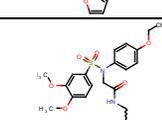 | -4.12 | 5.385 | 5.3820 | 0.003 |
| 11 | 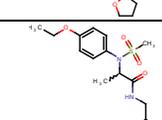 | -7.8 | 5.062 | 5.3820 | 0.32 |
| 12 | 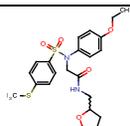 | -0.82 | 6.086 | 6.0867 | 0.0007 |
| 13 | 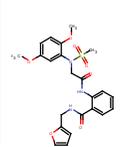 | 4.01 | 5.317 | 5.2282 | 0.0888 |
| 14 | 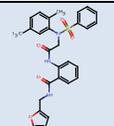 | 3.9 | 5.408 | 5.2282 | 0.1798 |
| 15 | 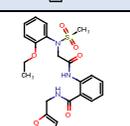 | -5.44 | 5.264 | 5.2282 | 0.0358 |
| 16 | 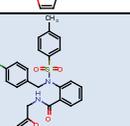 | 0.15 | 6.823 | 5.3820 | 1.441 |
| 17 | 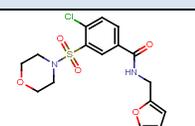 | 6.79 | 5.168 | 5.3820 | 0.214 |
| 18 | 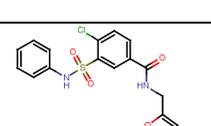 | 9.26 | 5.033 | 5.3820 | 0.349 |
| 19 | 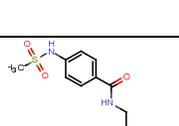 | -3.15 | 5.501 | 5.3820 | 0.119 |
| 20 | 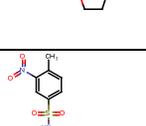 | 6.04 | 5.218 | 5.5359 | 0.3179 |
| 21 | 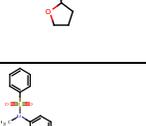 | -3.47 | 5.459 | 5.3820 | 0.077 |

| 22 | 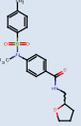 | -4.23 | 5.373 | 5.3820 | 0.009 |
|---|---|---|---|---|---|
| 23 | 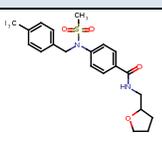 | -2.48 | 5.605 | 5.3820 | 0.223 |
| 24 | 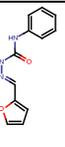 | -1.65 | 5.782 | 5.3820 | 0.4 |
| 25 | 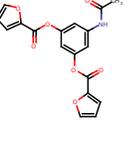 | -3.62 | 5.441 | 5.6625 | 0.2215 |
| 26 | 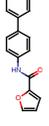 | 6.2 | 5.207 | 5.3820 | 0.175 |
| 27 | 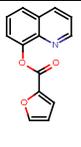 | 98.23 | 4.007 | 4.0090 | 0.002 |
| 28 | 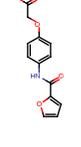 | -5.11 | 5.291 | 5.2282 | 0.0628 |
| 29 | 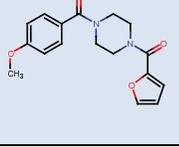 | 1.15 | 5.939 | 5.2282 | 0.7108 |
| 30 | 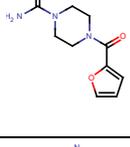 | -6.59 | 5.181 | 5.2282 | 0.0472 |
| 31 | 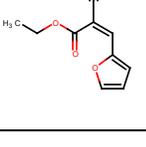 | -0.08 | 7.096 | 7.0986 | 0.0026 |

| 32 | 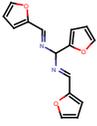 | 0.15 | 6.823 | 6.7122 | 0.1108 |
|----|----|----|----|----|----|
| 33 | 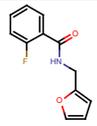 | -2.55 | 5.593 | 5.3820 | 0.211 |
| 34 | 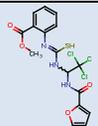 | -1.05 | 5.978 | 5.0744 | 0.9036 |
| 35 | 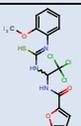 | -9.38 | 5.027 | 5.2282 | 0.2012 |
| 36 | 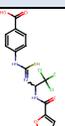 | -4.8 | 5.318 | 5.0744 | 0.2436 |
| 37 | 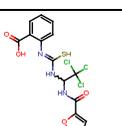 | 10.73 | 4.969 | 5.0744 | 0.1054 |
| 38 | 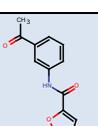 | 6.25 | 5.204 | 5.2282 | 0.0242 |
| 39 | 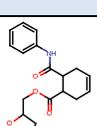 | 4.26 | 5.370 | 5.2282 | 0.1418 |
| 40 | 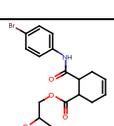 | -8.01 | 5.096 | 5.2282 | 0.1322 |

Table 1 Molecular structure with Observed  and Predicted activity of 5-nitrofuran-2-yl   used in training and test set using Model-1 (PLS)     Expt. = Experimental activity, Pred. = Predicted activity   IC50a = Compound concentration in micro mole required to inhibit growth by 50%

PIC50b = -Log (IC50 $\times$ $10^{-6}$): Training data set developed using model 1 (PLS) and Test set is light blue shaded.

## 2.2 BIOLOGICAL OBSERVED ACTIVITY DATA

For the evolution of QSAR models of 5-nitrofuran-2-yl ,of processes antitubercular activity in terms of half maximum inhibitory concentration IC50 (µM) versus (H37Rv) strains were took from the antitubercular drug molecule databases[12]. The IC50 activity data contains only molecules that have at least exhibited some activity. The biological activity data (IC50) were transformed in to pIC50 according to the formula pIC50 = $(-\log (IC50 \times 10^{-6}))$ was used as response values, thus correlating the data linear to the free energy change.

## 2.3 DESCRIPTOR CALCULATION FOR MOLECULAR DATASET

The VLife MDS tool used for the computation of various molecular descriptors containing topological index (J), connectivity index (x), radius of gyration (RG), moment of inertia, Wiener index(W), balabian index(J), centric index, hosoya index (Z), information based indices, XlogP, logP, hydrophobicity, elemental count, path count, chain count, pathcluster count, molecular connectivity index (chi), kappa values, electro topological state indices, electrostatic surface properties, dipole moment, polar surface area(PSA), alignment independent descriptor (AI)[11,14. The calculated molecular descriptors were gathered in a data matrix. The preprocessing for the generated molecular descriptors was done by removing invariable (constant column) and cross-correlated descriptors (with r = 0.99). which happen in total 156, 125 and 162 molecular descriptors for MLR, PCR and PLS accordingly to be used for QSAR analysis.

## 2.4 CREATION OF TRAINING AND TEST SET

The dataset of forty molecular descriptors is split s into training and test set by Sphere Exclusion (SE)[15-16] technique. In this technique initially data set splits into training and test set using sphere exclusion technique. In this technique variance value provides an idea to handle training and test set size. It needs to be adapted by trial and error until a desired split of training and test set is acquired. Increase in dissimilarity value results in increase in number of molecules in the test set. This technique is used for MLR, PCR and PLS models with pIC50 activity data as response variable and various 2D molecular descriptors computed for the molecules as independent variables.

## 2.5 MODEL VALIDATION

Model validation [17-18] is a essential manner of quantitative structure–activity relationship (QSAR) modelling. This is done to test the internal stability and predictive capability of the QSAR models. These three QSAR models were validated by the following method.

### 2.5.1 INTERNAL MODEL VALIDATION

Internal model validation was carried out using leave-one-out (LOO-$Q^2$) method. For calculating $q^2$, each sample in the training set was eliminated once and the activity of the eliminated sample was predicted by using the model developed by the remaining samples. The $Q^2$ computed using the expression which explains the internal strength of a model.

$$Q^2 = 1 - \frac{\sum(Y_{pred} - Y_{obs})^2}{\sum(Y_{obs} - Y_{mean})^2} \qquad (1)$$

In Eq. (1), $Y_{pred}$ and $Y_{obs}$ indicate predicted and observed activity values accordingly and $Y_{mean}$ signify mean activity value. A model is considered acceptable when the value of $Q^2$ exceeds 0.5.

## 2.5.2 EXTERNAL MODEL VALIDATION

External model validation, the activity of each sample in the test set was predicted using the model created by the training set. The pred_$r^2$ value is computed as follows.

$$\text{pred\_}r^2 = \frac{\sum(Y_{pred(test)} - Y_{test})^2}{\sum(Y_{train} - Y_{mean(train)})^2} \qquad (2)$$

In Eq (2) $Y_{pred(test)}$ and $Y_{test}$ indicate predicted and observed activity values for the test set and $Y_{train}$ indicates mean activity value of the training set. For the predictive QSAR model, the value of pred_$r^2$ must be more than 0.5.

## 2.5.3 RANDOMIZATION TEST

Randomization test or Y-scrambling is key mean of statistical validation. To assess the statistical importance of the QSAR model for the dataset, one tail hypothesis testing is used. The strength of the models for training sets was tested by examining these models to those derived for random datasets. Random sets were produced by rearranging the activities of the samples in the training set. The statistical model was determined using different randomly reorganize activities (random sets) with the chosen molecular descriptors and the equivalent $Q^2$ were computed. The importance of the models for that reason obtained was developed based on a computed $Z_{score}$.

A Z score value is calculated by the following equation:

$$Z_{score} = \frac{(h - \mu)}{\sigma} \qquad (3)$$

Where $h$ is the $Q^2$ value computed for the dataset, $\mu$ the mean $Q^2$, and is its $\sigma$ standard deviation calculated for various iterations using models build by different random datasets. The probability (a) of importance of randomization test is derived by comparing $Z_{score}$ value with $Z_{score}$ critical value as stated, if $Z_{score}$ value is less than 4.0; otherwise it is computed by the expression as given in the literature. For example, a $Z_{score}$ value more than 3.10 proposes that there is a probability (a) of smaller than 0.001 that the QSAR model build for the dataset is random. The randomization test proposes that all the created models have a probability of less than 1% that the model is produced by chance.

## 2.6 MULTIPLE LINEAR REGRESSION (MLR) ANALYSIS

MLR technique used for modelling linear relationship between a response variable Y (pIC50) and independent variables X (2D molecular descriptors). MLR is based on least squares technique: the model is fit such that sum-of-squares of differences of actual and a predicted values are minimized. MLR estimates the regression coefficients ($r^2$) by applying least squares fitting

54

technique. The model builds a relationship in the form of a straight line (linear) that best estimates all the individual data points. In regression analysis, conditional mean of response variable (pIC50) Y depends on (molecular descriptors)X. MLR analysis add to this idea to include more than one independent variables. Regression expression takes the form.

$$Y = b_1x_1 + b_2x_2 + b_3x_3 + c \qquad (4)$$

where Y is a response variable, 'b's are regression coefficients for corresponding 'x's are molecular descriptors(independent variables), 'c' is a regression constant or intercept [19,25].

## 2.7 PRINCIPAL COMPONENT REGRESSION (PCR)  ANALYSIS

Principal Component Regression (PCR) is a regression technique that uses principal component analysis(PCA) when evaluating regression coefficients. PCR presents a technique for finding structure in datasets. Its object is to group correlated variables, replacing the earlier  descriptors by new set called principal components (PCs). These PC's are uncorrelated and are developed as a simple linear aggregation of earlier variables. It moves the data into a new set of axes such that first few axes indicates most of the variations within the data. First PC (PC1) is expressed in the direction of maximum variance of the whole dataset. Second PC (PC2) is the direction that defines the maximum variance in orthogonal subspace to PC1. Consequent components are taken orthogonal to the particular formerly chosen and defines best of remaining variance, by locating the data on new set of axes, it can points major fundamental structures certainly. Value of each point, when moved  to a given axis, is called the PC value. PCA chooses a new set of axes for the data. These are chosen in decreasing order of variance within the data. The aim of principal component PCR is the computation of values of a response variable on the basis of chosen PCs of independent variables [21].

## 2.8  PARTIAL LEAST SQUARES (PLS) REGRESSION  ANALYSIS

 PLS  is a well known regression technique  which can be used to correlate one or more response variable (Y) to various independent variables(X)  . PLS relates a matrix Y of response variables to a matrix X of molecular descriptors. PLS is useful in conditions where the number of molecular descriptors( independent variables) exceeds the number of samples, when X data contain colinearties or when N is less than 5M, where N is number of samples and M is number of response variables. PLS builds orthogonal components using existing correlations between independent variables and corresponding outputs  while also keeping most of the variance of independent variables. Major aim of PLS regression is to predict the activity (Y) from X and to define their common frame[22,23] . PLS is probably the least contrary of various multivariate extensions of MLR model. PLS is a technique for constructing predictive models when factors are many and highly collinear.

## 2.9 EVALUATION OF  THE QSAR MODELS

The created QSAR models are computed using the following statistical parameters: N (Number of samples in regression); K  (Number of independent variables(molecular descriptors)); DF (Degree of freedom); optimum component ( number of optimums); $r^2$ ( the squared correlation coefficient); F test (Fischer's Value)  for statistical importance; $q^2$  (cross-validated correlation coefficient); pred_r$^2$ ( r$^2$ for external test set); Z$_{score}$ ( Z score computed by the randomization test); Best_ran_r$^2$ (maximal r$^2$ value in the randomization test) ; Best_ran_q$^2$ (maximal q$^2$ value in the randomization test) ; α ( statistical importance parameter obtained by the randomization test). The correlation coefficient r$^2$ is a respective standard of fit by the regression expression. It expressed the part of the variation in the observed data is analyzed by the regression. Despite, a QSAR models are examined to be predictive, if the following prerequisites are satisfied: r$^2$ > 0.6 , q$^2$ > 0.6 and pred_r$^2$ > 0.5 [24] . The F-test indicates the ratio of variance described by the model and variance due to the error in the regression. High values of the F-test indicate that model is statistically meaningful. The reduced standard error of pred_r$^2$se , q$^2$_se and r$^2$_se demonstrates actual value of the fitness of the model. The cross-correlation extent was set at 0.5.

# 3. RESULTS

Taining set of 30 and 10 of test set of 5-nitrofuran-2-yl having different substitution were employed.

## 3.1 CREATION OF QSAR MODELS

### 3.1.1 PARTIAL LEAST SQUARES (PLS) REGRESSION ANALYSIS

The molecular descriptors were applied to PLS technique to developQSAR models by using simulated anealing variable selection mode. PLS model is having following QSAR Eq.(5) with five descriptors.

$$pIC50 = 1.8704(StsCcount) + 4.0747(chi5chain) - 0.6865(SaaaCcount) + 0.7046(SssScount) - 0.1538(SdssCcount) + 4.9478 \tag{5}$$

Table 2 Statistical parameters of PLS, MLR And PCR

| Parameters | PLS | MLR | PCR |
|---|---|---|---|
| N | 40 | 40 | 40 |
| DF | 26 | 24 | 28 |
| r$^2$ | 0.8484 | 0.8484 | 0.3289 |
| q$^2$ | 0.0939 | 0.0932 | −5.3805 |
| F-test | 48.5187 | 26.8725 | 13.7231 |
| best_ran_r$^2$ | 0.56429 | 0.28620 | 0.32891 |
| best_ran_q$^2$ | −0.03892 | −0.08598 | −2.93782 |
| Z$_{score\_ran\_r}$$^2$ | 3.43122 | 8.11471 | 2.81258 |
| Z$_{score\_ran\_q}$$^2$ | 1.59111 | 1.31886 | −2.47533 |
| α_ran_r$^2$ | 0.00100 | 0.00000 | 0.01000 |
| α_ran_q$^2$ | 0.10000 | 0.10000 | 99.00000 |
| r$^2$_se | 0.2277 | 0.2370 | 0.4617 |
| q$^2$_se | 0.5568 | 0.5797 | 1.4237 |
| pred_r$^2$ | −0.5604 | −0.5616 | −0.0734 |
| pred_r$^2$se | 0.7252 | 0.7255 | 0.6015 |

The above analysis directs to the improvement of statistically meaningful QSAR model, which allows understanding of the molecular properties/features that play an key role in governing the variation in the activities. In addition, this QSAR study allowed examining influence of very simple and easy-to-compute molecular descriptors in discovering biological activities, which could shed light on the important factors that may aid in design of new potent molecules.

All the parameters and their significance, which contributed to the specific antitubercular inhibitory activity in the generated model is discussed below.

**1. StsCcount:** This descriptor indicates the total number of carbon atoms with a triple bond and a single bond exist in the molecule. Positive Contribution of this descriptor to the model is 31.72%.

**2.chi5chain:** This descriptor signifies a retention index for five membered ring. Positive Contribution of this descriptor to the model is 21.99%.

**3. SaaaCcount:** This descriptor defines the total number of carbon connected with three aromatic bonds. Negative Contribution of this descriptor to the model is -23.28%.

**4. SssSount:** This descriptor indicates the total number of sulphur atom attached with two single bonds. Positive Contributions of this descriptor to the model is 11.95%.

**5. SdssCcount:** This descriptor defines the total number of carbon connected with one double and two single bond. Negative Contribution of this descriptor to the model is -11.06%.

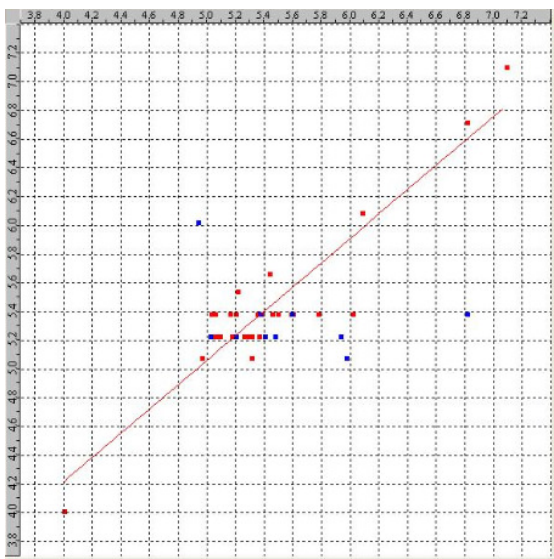

Figure 1 Observed vs. Predicted activities for training and test set molecular descriptors by Partial Least Square model. (A) Training set (Red dots) (B) Test Set (Blue dots).

The PLS model gave correlation coefficient ( $r^2$ ) of 0.8484, significant cross validated correlation coefficient ( $q^2$ ) of 0.0939, F-test of 48.5187 and degree of freedom 26. The model is validated by α_ran_$r^2$= 0.00100, α_ran_$q^2$ = 0.10000, best_ran_$r^2$ = 0.56429, best_ran_$q^2$= -0.03892, $Z_{score\_ran\_r}^2$= 3.43122 and $Z_{score\_ran\_q}^2$ =1.59111. The randomization test proposes that the created model have a probability of smaller than 1% that the model is build by chance. Statistical data is presented in Table 2. The graph of observed vs. predicted activity is demonstrated in Figure 1. The descriptors which contribute for the QSAR model is demonstrated in Figure 2.

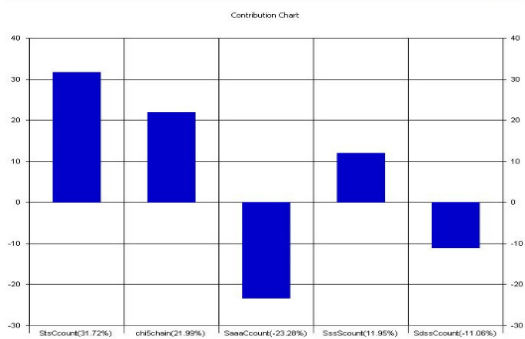

Figure 2 Percentage contribution of each descriptor in created PLS model describing variation in the activity

### 3.1.2 MULTIPLE LINEAR REGRESSION (MLR) ANALYSIS

The QSAR analysis by Multiple Linear Regression method with simulated annealing variable selection technique, the final QSAR model is created having five descriptors is shown in Eq. (6).

$$pIC50 =$$
$$1.8681(\pm 0.2421)StsCcount + 4.0722(\pm 0.7599)chi5chain -$$
$$0.6879(\pm 0.1139)SaaaCcount + 0.7033(\pm 0.2393)SssSscount -$$
$$0.1548 (\pm 0.0265)SdssCcount + 4.9497 \quad\quad (6)$$

MLR Model has a correlation coefficient ($r^2$) of 0.8484, significant cross validated correlation coefficient ($q^2$) of 0.0932, F test of 26.8725 and degree of freedom 24. The model is validated by $\alpha\_ran\_r^2 = 0.00000$, $\alpha\_ran\_q^2 = 0.10000$, $best\_ran\_r^2 = 0.28620$, $best\_ran\_q^2 = -0.08598$, $Z_{score\_ran\_r}^2 = 8.11471$ and $Z_{score\_ran\_q}^2 = 1.31886$

The randomization test proposes that the created model have a probability of smaller than 1% that the model is build by chance. The observed and predicted values with residual values are demonstrated in Table 1.Statistical data is demonstrated in Table 2.The graph of observed vs. predicted activity demonstrated is in Figure 3. The descriptors which contribute for the QSAR model are demonstrated in Figure 4. All the parameters and their significance, which contributed to the specific antitubercular inhibitory activity in the generated models are explained below.

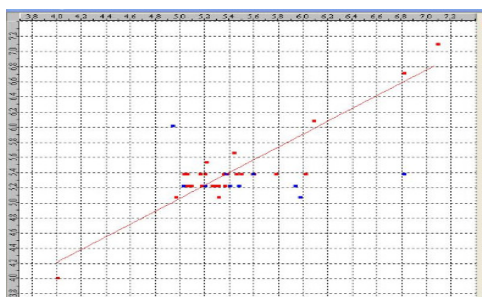

Figure 3 Observed vs. Predicted activities for training and test set molecular descriptors from the Multiple Linear Regression model. (A) Training set (Red dots) (B) Test Set (Blue dots).

**1. StsCcount:** This descriptor indicates the total number of carbon atoms with a triple bond and a single bond present in the molecule. Positive Contribution of this descriptor to the model is 31.67%.

**2.chi5chain:** This descriptor signifies a retention index for five membered ring. Positive Contribution of this descriptor to the model is 21.97%.

**3. SaaaCcount:** This descriptor defines the total number of carbon connected with three aromatic bonds. Negative Contribution of this descriptor to the model is -23.32%.

**4. SssScount:** This descriptor indicates the total number of sulphur atom attached with two single bonds. Positive Contributions of this descriptor to the model is 11.92%.

**5. SdssCcount:** This descriptor defines the total number of carbon connected with one double and two single bond. Negative Contribution of this descriptor to the model is -11.12%.

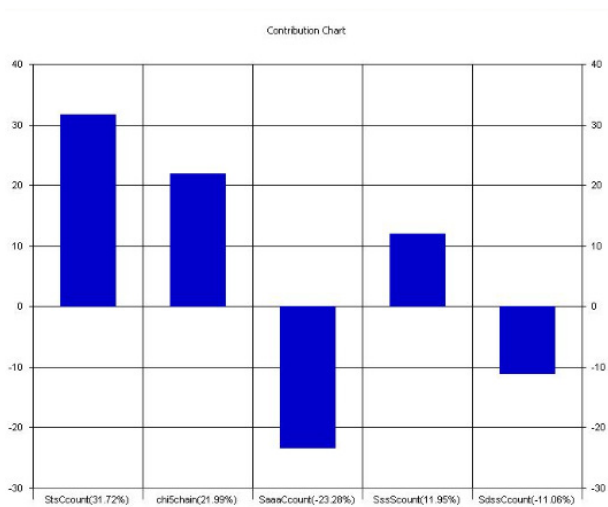

Figure 4  Percentage contribution of each descriptor in created MLR model describing variation in the activity.

### 3.1.3  PRINCIPAL COMPONENT REGRESSION (PCR) ANALYSIS

The molecular descriptors were applied to under goes PCR technique to create QSAR model with Simulated anealining variable selection mode by using PCR model. The final QSAR  model is Eq. (7) was created having one descriptor as follows.

$$pIC50 = 1.7397 StsCcount + 5.3563 \qquad (7)$$

The PCR model gave correlation coefficient ($r^2$)  is 0.3289, significant cross validated correlation coefficient ($q^2$) of -5.3805, F test of 13.7231 and degree of freedom 28. The model is validated by $\alpha\_ran\_r^2 = 0.01000$, $\alpha\_ran\_q^2 = 99.00000$, best$\_ran\_r^2 = 0.32891$, best$\_ran\_q^2 = -0.13938$ , $Z_{score\_ran\_r}^2 = 2.81258$ and $Z_{score\_ran\_q}^2 = -2.47533$. The randomization test proposes that the created model have a probability of smaller than 1% that the model is build by chance.

Statistical data is demonstrated in Table 2. The graph of observed vs. predicted activity is in demonstrated Figure 5 .The descriptors which contribute for the QSAR model is demonstrated in Figure 6.

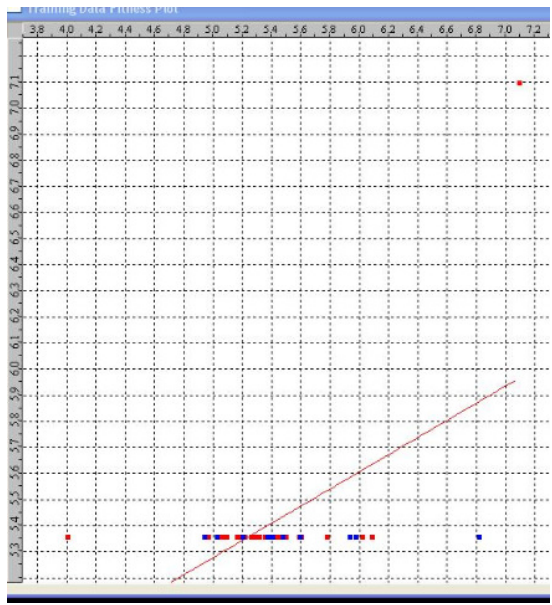

Figure 5  Observed vs. Predicted activities for training and test set molecular descriptors by Principal Component Regression model. A) Training set (Red dots) B) Test Set (Blue dots).

All the parameters and their significance, which contributed to the specific   antitubercular inhibitory activity in the generated models are discussed here.

**1. StsCcount:** This descriptor indicates the total number of  carbon atoms with a triple bond and a single bond    present in the molecule. Positive Contribution of this descriptor to the model is 100%.

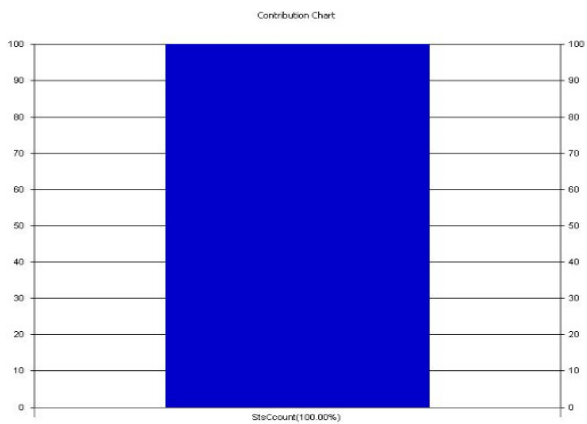

Figure 6 Percentage contribution of each descriptor in developed PCR model describing variation in the activity

# 4. CONCLUSION

The 2D QSAR analysis were conducted with a series of 5-nitrofuran-2-yl derivatives for mycobacterium tuberculosis(H37Rv) inhibitors , and some useful  predictive  models were obtained. The physicochemical  molecular descriptors were found to have an key role in governing the change in activity. The statistical parameters demonstrate the estimation power of QSAR model for the molecular descriptor data set from which it has been determined and evaluate it only internally. The overall performance of prediction was found to be around 84% in case of PLS and MLR. Among the three 2D-QSAR models (MLR, PCR, and PLS), the results of PLS and MLR analysis showed significant predictive power and reliability as compare to PCR technique.

## ACKNOWLEDGEMENTS

The Authors are thankful to Dr Mahesh .B. Palkar Department of Pharmaceutical Chemistry K.L.E Pharmacy College Hubli

## Authors

**Doreswamy** received B.Sc degree in Computer Science and   M.Sc Degree in Computer Science from  University of  Mysore in 1993 and 1995 respectively. Ph.D degree in Computer Science from   Mangalore University in the year 2007. After completion of his Post-Graduation Degree, he subsequently joined and served asLecturer in Computer Science at St. Joseph's College, Bangalore from 1996 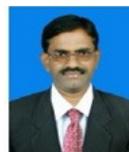 1999.Then he has   elevated to the position  Reader in Computer Science at  Mangalor Universityin year 2003. He was the Chairman of the Department of  Post-Graduate Studies  and research in computer science from 2003-2005 and from 2009-2008 and served at varies capacitiesin Mangalore University at present he is the Chairman  of Board of Studies and Professor  in Computer Science of Mangalore University. His areas of  Research  interests include Data Mining and Knowledge Discovery, Artificial Intelligence and Expert Systems, Bioinformatics ,Molecular modelling and simulation ,Computational Intelligence ,Nanotechnology, Image Processing and  Pattern  recognition. He has  been granted a Major   Research   project entitled "Scientific Knowledge Discovery Systems (SKDS) for Advanced Engineering Materials Design Applications" from the funding   agency University Grant Commission, New Delhi , India. He  has been  published about 30 contributed  peer  reviewed Papers at national/International Journal  and  Conferences. He received SHIKSHA RATTAN PURASKAR for his outstanding   achievements  in  the  year 2009 and RASTRIYA VIDYA SARASWATHI AWARD for outstanding  achievement in chosen  field of  activity  in the  year  2010.

**Chanabasayya.M.Vastrad** received   B.E. degree and M.Tech. degree in the year 2001and 2006   respectively. Currently working towards his Ph.D Degree in Computer Scienceand  Technology  under the guidance of  Dr. Doreswamy in the Department  of Post-Graduate Studies  and  Research  in  Computer Science , Mangalore University. 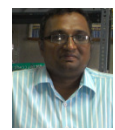